\documentclass{aastex63}
\usepackage{graphicx}
\usepackage{amssymb}                    
\usepackage{ulem}
\usepackage{float}
\usepackage{rotating}
\usepackage{enumitem}
\usepackage{rotating}
\usepackage{makecell}
\usepackage{appendix}
\usepackage{fixmath}
\usepackage{color}

\newcommand{\x}{\ensuremath{\times}}

\usepackage{times} 
\usepackage{amsmath}
\submitjournal{ApJ}

\shortauthors{Athiray et al.}
\shorttitle{{\magixs} Calibration I : X-ray source}

\newcommand{\magixs}{\textit{MaGIXS}}

\begin{document}


\title{Calibration of the {\magixs} experiment I: Calibration of the X-ray source at the X-ray and Cryogenic Facility (XRCF)}

\correspondingauthor{P.S. Athiray}
\email{athiray.panchap@nasa.gov}

\author[0000-0002-4454-147X]{P.S. Athiray}
\affil{NASA Postdoctoral Program, NASA Marshall Space Flight Center, Huntsville, AL 35812}

\author[0000-0002-5608-531X]{Amy R.\ Winebarger}
\affil{NASA Marshall Space Flight Center, Huntsville, AL 35812}

\author[0000-0002-7139-6191]{Patrick Champey}
\affil{NASA Marshall Space Flight Center, Huntsville, AL 35812}

\author{Ken Kobayashi}
\affil{NASA Marshall Space Flight Center, Huntsville, AL 35812}

\author[0000-0002-7219-1526]{Genevieve D.\ Vigil}
\affil{NASA Postdoctoral Program, NASA Marshall Space Flight Center, Huntsville, AL 35812}


\author{Harlan Haight}
\affil{NASA Marshall Space Flight Center, Huntsville, AL 35812}

\author{Steven Johnson}
\affil{NASA Marshall Space Flight Center, Huntsville, AL 35812}

\author[0000-0001-9076-6461]{Christian Bethge}
\affil{Universities Space Research Association, NASA Marshall Space Flight Center, Huntsville, AL 35812}

\author[0000-0002-3770-009X]{Laurel A. Rachmeler}
\affil{NASA Marshall Space Flight Center, Huntsville, AL 35812}
\affil{National Centers for Environmental Information (NOAA), Boulder, CO 80305}

\author[0000-0002-6172-0517]{Sabrina Savage}
\affil{NASA Marshall Space Flight Center, Huntsville, AL 35812}

\author{Brent Beabout}
\affil{NASA Marshall Space Flight Center, Huntsville, AL 35812}

\author{Dyana Beabout}
\affil{NASA Marshall Space Flight Center, Huntsville, AL 35812}

\author{William Hogue}
\affil{NASA Marshall Space Flight Center, Huntsville, AL 35812}

\author{Anthony Guillory}
\affil{NASA Marshall Space Flight Center, Huntsville, AL 35812}

\author{Richard Siler}
\affil{NASA Marshall Space Flight Center, Huntsville, AL 35812}

\author{Ernest Wright}
\affil{NASA Marshall Space Flight Center, Huntsville, AL 35812}

\author{Jeffrey Kegley}
\affil{NASA Marshall Space Flight Center, Huntsville, AL 35812}

\begin{abstract}
The Marshall Grazing Incidence Spectrometer (\magixs) is a sounding rocket experiment that will observe the soft X-ray spectrum of the Sun from 24 - 6.0 \AA\ (0.5 - 2.0 keV) and is scheduled for launch in 2021. Component and instrument level calibrations for the {\magixs} instrument are carried out using the X-ray and Cryogenic Facility (XRCF) at NASA Marshall Space Flight Center.  In this paper, we present the calibration of the incident X-ray flux from the electron impact source with different targets at the XRCF using a CCD camera; the photon flux at the CCD was low enough to enable its use as a ``photon counter'' i.e. the ability to identify individual photon hits and calculate their energy. The goal of this paper is two-fold: 1) to confirm that the flux measured by the XRCF beam normalization detectors is consistent with the values reported in the literature and therefore reliable for {\magixs} calibration and 2) to develop a method of counting photons in CCD images that best captures their number and energy.  
\end{abstract}



\section{Introduction} \label{sec:intro}

For over four decades, X-ray, EUV, and UV spectral observations have been used to measure physical properties of the solar atmosphere. During this time, there has been substantial improvement in the spectral, spatial, and temporal resolution of the observations in the EUV and UV wavelength ranges. At wavelengths below 100\AA, however, observations of the solar corona with simultaneous spatial and spectral resolution are limited, and not since the late 1970's have spatially resolved solar X-ray spectra been measured. Because the soft X-ray regime is dominated by emission lines formed at high temperatures, X-ray spectroscopic techniques yield insights to fundamental physical processes that are not accessible by any other means.

The Marshall Grazing Incidence X-ray Spectrometer (\magixs) will measure, for the first time, the solar spectrum from 6-24\AA\, with a spatially resolved component along an $>8$\arcmin\ slit \citep{kobayashi2010,kobayashi2018, champey2016}. {\magixs} will provide a good diagnostic capability for high-temperature, low-emission measure plasma, which will help to determine the frequency of heating events in coronal structures \citep{Athiray2019}. The instrument, comprised of a Wolter-I telescope mirror, a grazing incidence spectrograph, and a CCD detector, was aligned using a UV centroiding instrument designed for the \textit{Chandra X-ray Observatory} \citep{Glen1995} and a visible-light auto-collimating theodolite. Several small reference mirrors and alignment reticles were co-aligned to the optical axis of the grazing incidence mirrors during the mounting process. These references were used later to co-align the telescope mirror with the spectrometer assembly (Champey et al., in preparation). A series of X-ray tests were performed iteratively during the alignment process to confirm alignment, focus the telescope on the slit, and measure throughput. X-ray testing of the mounted telescope mirror was performed in the Stray Light Facility (SLF) \citep{champey2019}, while the integrated instrument was tested in the X-ray and Cryogenic Facility (XRCF) at NASA Marshall Space Flight Center. The remaining set of end-to-end X-ray tests are designed to calibrate the spectrograph dispersion (wavelength calibration) and measure throughput of \magixs. These tests will use the XRCF's electron impact point source (EIPS) and the four targets listed in Table~\ref{tab:calib_targets}.

Though absolute radiometric calibration is not required for this sounding rocket instrument, the data gathered at the XRCF make radiometric calibration possible if the beam normalization detectors (BND), available at the XRCF, can be used to provide an accurate measure of the photon flux of the X-ray beam.  Because these detectors have not been calibrated since 1998 \citep{Martin1997}, we performed an experiment to verify the BND performance, as well as develop and verify our method to detect photons and calculate their energies in a CCD detector identical to the {\magixs} X-ray detector.   

We present our experiments, methods of analysis, and results. In Section \ref{sec:setup} we review the experiment details and data collection. In Section \ref{sec:eventselection} we discuss the Monte Carlo simulation used to develop the event selection algorithm.  In Section \ref{sec:results}, we present the results of applying this method to the test data.  We find good agreement (to within 20\%) between the photon flux measured by the BND and the CCD.  In a follow-on paper, we will apply these techniques to the \magixs\ instrument, tested in different configurations, determine the component level radiometry, and predict the solar signal for typical active regions.

\section{Experiments}
\label{sec:setup}

\subsection{Setup}
\begin{figure}
    \centering
    \includegraphics[width=0.75\linewidth]{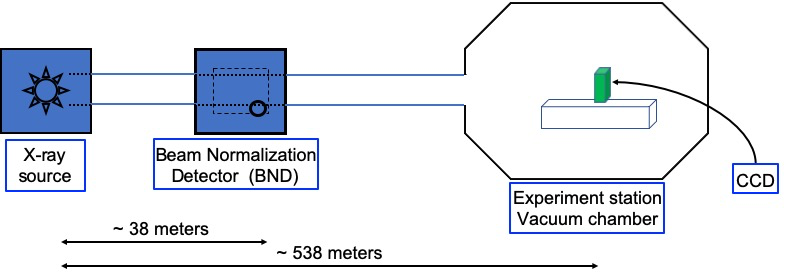}
    \caption{Schematic representation of experiment setup at the X-ray and Cryogenic Facility at NASA Marshall.}
    \label{fig:setup}
\end{figure}

A schematic representation of the experimental setup is shown in Figure~\ref{fig:setup}. The test setup includes an X-ray source (EIPS), the BND, and a CCD with back-end readout electronics and a liquid nitrogen (LN2) cooling system. The EIPS has selectable targets and a filter wheel populated with continuum reducing filters of nominally 2~$\times$ mean free path thickness at the desired spectral line. The X-ray beam from the source is transmitted through an evacuated, 518 m long guide tube to an evacuated 7.3 m diameter $\times$ 22.9 m long instrument chamber. The source is operated with an anode voltage that is 4~$\times$ the L-shell or 5~$\times$ the K-shell binding potential of the target at a permissible current. A list of the calibration targets, along with corresponding line energies useful for {\magixs} calibration, is given in Table~\ref{tab:calib_targets}. The targets employed in the EIPS are the same ones used in the calibration of Chandra X-ray mirrors, except for target Mg, which has been replaced with a new Mg target.  The BND is mounted on a translation stage within the guide tube at a distance of 38 meters from the source, and outside of the EIPS-CCD beam path. The BND is preceded by a positionable masking plate with apertures of calibrated diameters (see Table~\ref{tab:calib_targets}), used to avoid saturation of the detector. The CCD, along with readout avionics and cooling hardware, are mounted on a translation stage and are placed within the instrument chamber, 538 m  from the source.

{}

\begin{deluxetable}{c c c c c c c}
\tablecaption{List of calibration targets and respective line energies. \label{tab:calib_targets}}
\tablehead{
\colhead{S.No}&\colhead{Target} &\colhead{Line energy}& \colhead{Voltage}& \colhead{Current}& \colhead{Filter} & \colhead{BND aperture diameter} \\
\colhead{}&\colhead{}  &\colhead{keV}& \colhead{kV}& \colhead{mA} &\colhead{}&\colhead{cm}}
\startdata
1&Ni-L& 0.85& 3.40 & 5.8 & Ni-L2 &0.4 \\
2&Cu-L & 0.93& 3.70 & 7.7 & Cu-L2&0.4\\
3&Zn-L & 1.01& 4.00 & 3.0 &No filter &0.1\\
4&Mg-K & 1.25& 6.50 & 1.5 &Mg-K2&0.4\\
\enddata
\end{deluxetable}
\subsection{Data collection and pre-processing}
The BND is a flow proportional counter (FPC) using flowing P10 gas (90\% Argon, 10\% Methane) at a pressure of 400 Torr. It has a 15{\arcsec}~$\times$~5{\arcsec} Aluminum-coated polyimide window supported by a gold-coated tungsten wire grid. The FPC anode voltage controls its avalanche gain, and can be set to a voltage appropriate for the energy of the line being monitored. A signal processing chain follows the FPC, consisting of a preamplifier, shaping amplifier, and multi-channel buffer analog to digital converter. The data collected are transferred to a computer for archival and analysis. For a complete description of the FPC, see \citet{Bradford1997}.
 
The CCD employed for this test is a back-illuminated, ultra-thinned, astro-processed sensor with an active area that contains 2K~$\times$~1K pixels with 15 $\mu$m pixel size, similar to the {\magixs} flight camera. The CCD is mounted in a copper carrier that is connected via a copper strap to a cold block. The cold block is actively cooled to roughly -100$^{\circ}$\,C using liquid nitrogen, which results in a CCD carrier average temperature of roughly -70$^{\circ}$\,C.  The carrier temperature is controlled such that it is maintained constant within $\pm$~5$^{\circ}$\,C. 

\begin{figure}
    \centering
     \includegraphics[width=0.5\linewidth]{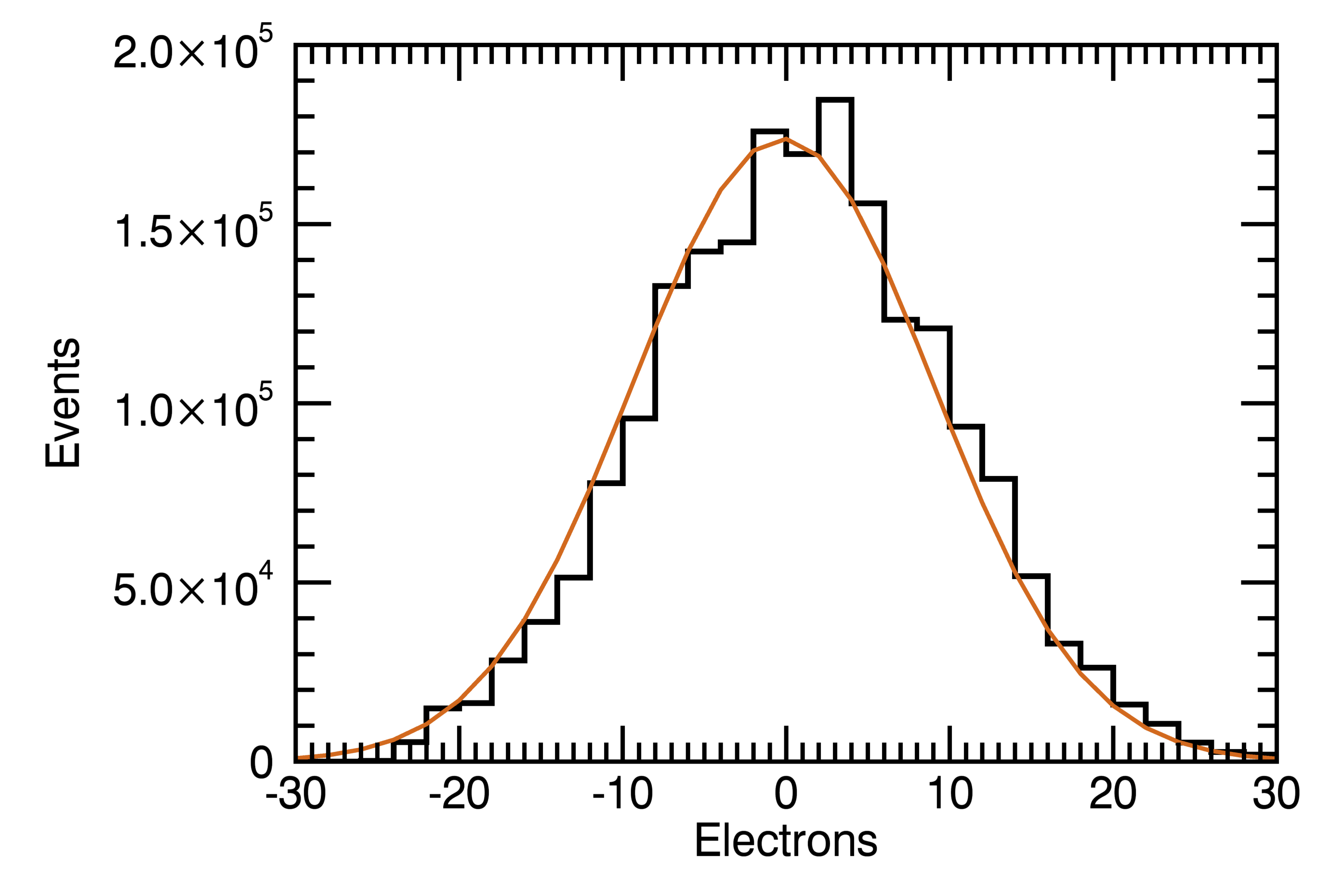}
    \caption{Histogram of the residuals from the processed data (bias, dark current, and fixed pattern noise subtracted) demonstrates that the measured RMS read noise is $\sim$ 9\,e$^-$.}
    \label{fig:Noise_spectrum}
\end{figure}

The low-noise camera has been developed by NASA Marshall Spaceflight Center as one of a series of cameras developed for suborbital missions \citep{Laurel2019, champey2014, Champey2015}. The CCD is operated in frame transfer mode, with 1k~$\times~$2k pixels exposed and two 500~$\times$~2k readout regions that are mechanically masked.  There are four read-out taps on the detector; it requires roughly 1.2\,s to read out 1k$\times$2k pixels. All exposures are 2\,s long, meaning the CCD is operating with $\sim$100\% duty cycle.  More than 200 frames are collected for every target in Table~\ref{tab:calib_targets} for adequate statistics. Additionally, during each data run, dark images are acquired by closing a gate valve between the source and CCD detector.   The raw CCD images are processed using standard reduction procedures to remove bias level, dark current, and fixed pattern noise in the images. The gain of the camera was determined prior to XRCF testing using the Mn K-$\alpha$ and K-$\beta$ lines from a sealed radioactive source Fe$^{55}$ and found to be $\sim$ 2.58 electrons per data number (DN). The variability of gain between the quadrants are less than 0.03 electron DN$^{-1}$, which means the systematic  uncertainty introduced from gain is less than $\sim$ 2\%. The images are converted from Data Number (DN) to electrons using the known gain of the camera. A histogram of the residual distribution indicates the RMS read noise is approximately 9\,e$^-$ as shown in Figure~\ref{fig:Noise_spectrum}.

If the photon flux is low and the energy of the photons are much larger than the noise in the detector, we can detect individual photons and measure their associated energy.  These conditions are satisfied by controlling the source strength with the adjustment of current/voltage settings. Appropriate filters are used at the source end to reduce the low energy bremsstrahlung continuum and lines arising from any anode contaminants. The optimal settings used for our experiment are listed in Table~\ref{tab:calib_targets} (Column 3, 4 and 5). We mention that during the experiment, no filter was placed when we operated target Zn, which was unintentional.

\section{Event recognition}
\label{sec:eventselection}

In order to successfully perform photon counting using CCD, events registered from photon hits are to be selected properly and eliminate all other sources of noise that are present in CCD. One of the challenges with the CCD is the relatively small pixel size (15 $\mu$m). The charge clouds produced from the absorbed X-ray photons drift under the electric field and undergo lateral diffusion. As a result, charge induced by the X-ray events may be shared over many pixels, known as {\it charge shared events}. To calculate the energy of a photon, it is important that the charge collected in multiple pixels are measured accurately and added with appropriate selection criteria to precisely determine the incident photon flux. Event reconstruction demands a good understanding of detector noise and uniformity in the pixel response. Typically thresholds are applied to exclude noisy outliers and maximize event recovery from multipixel events. Our goal is to develop a method and demonstrate how well we can determine the incident photon flux from CCD images by maximizing the event recovery from multipixel events.

 {}

\subsection{Simulation}
To understand the effect of pixel size and detector noise, and to optimize event reconstruction from charge shared events, we performed Monte-Carlo simulations of photon interaction and charge propagation in the CCD, which is assumed to have properties similar to the {\magixs} detector. The model assumes a simple 1D electric field obtained by solving Poisson's equation. Some of the basic assumptions and calculations for the electric field are obtained from \citet{Athiray2015}. We mention that our model is simplified with many assumptions and does not simulate charge loss mechanisms. We consider this `toy' model as an approximation to simulate multipixel events by predicting the response to incident X-ray photons in a CCD.  This tool provides a method with which to test and validate the event selection method, which can be further fine-tuned for `real' experiment data.  The physical device parameters of the CCD used in the simulation are listed in Table~\ref{tab:params}. 

\begin{figure}
    \includegraphics[width=0.5\linewidth]{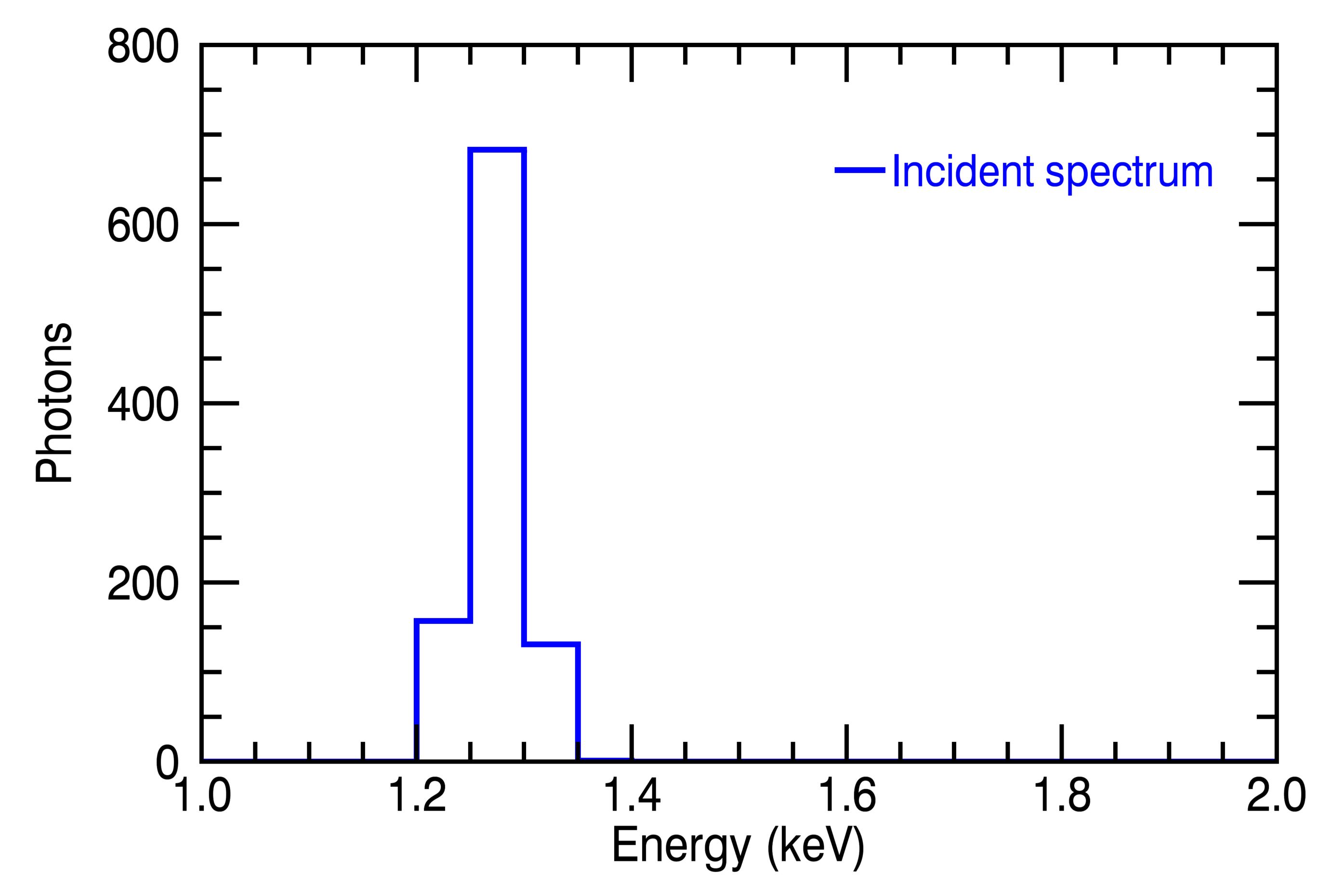}
    \includegraphics[width=0.5\linewidth]{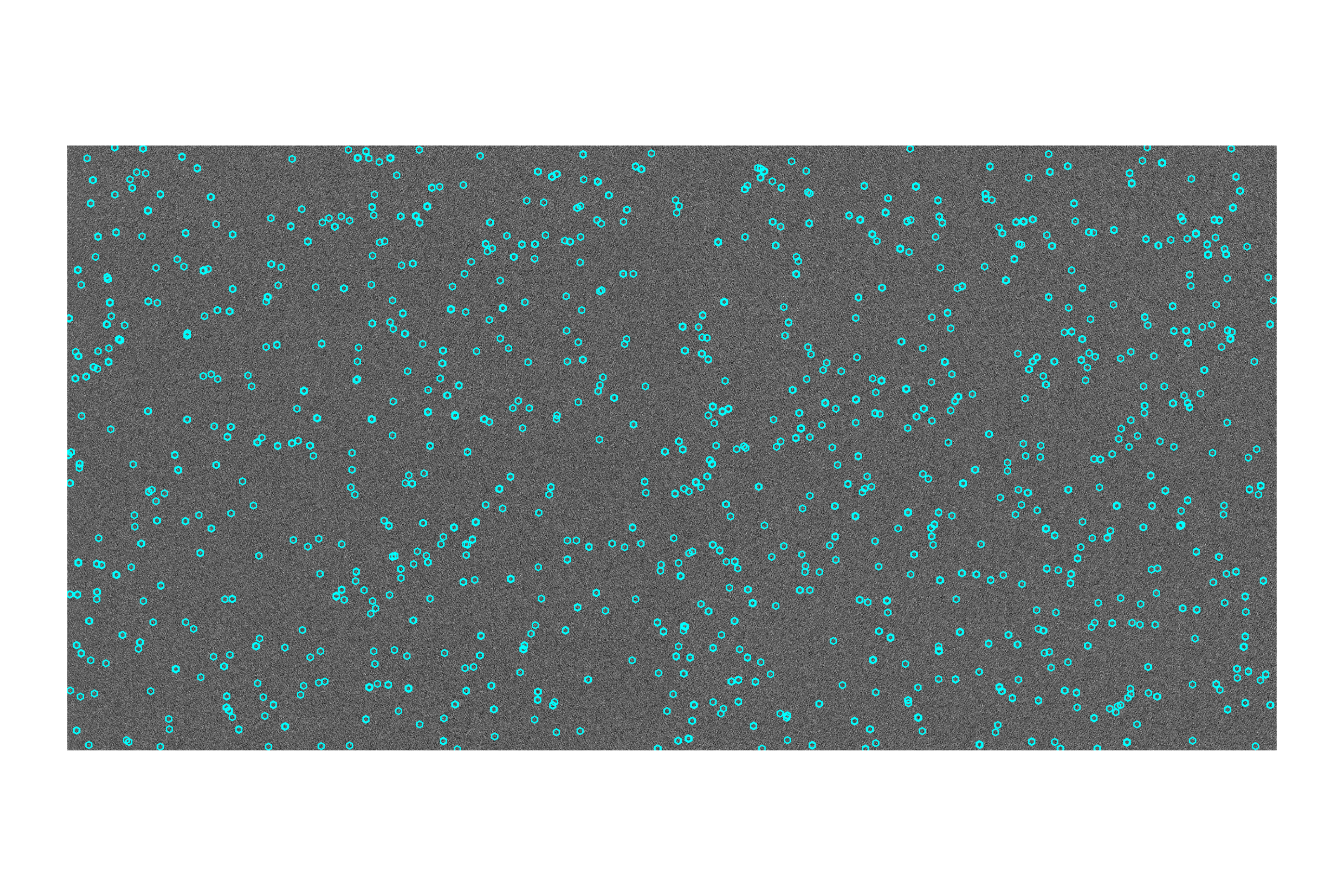}
    \caption{(Left) The incident photon spectrum with a mean energy at 1.25\,keV incorporated with Fano noise using the Fano factor F = 0.115 simulated to enter the CCD. (Right) Synthetic X-ray image showing simulated photon hits (Cyan) with mean energy of 1.25\,keV combined with a noise spectrum shown in Figure~\ref{fig:Noise_spectrum}.}
    \label{fig:source_noise_spectrum}
\end{figure}

The variance in the production of number of electron-hole per X-ray photon will always be less than the Poisson statistical variance, which is known as the Fano noise. The Fano noise determines the inherent line width of an X-ray detector. To incorporate this effect, we simulated the incident spectrum with a mean energy E$_i$ along with the Fano noise using:
\begin{equation}
E_f=E_i + R_n(0) \sqrt{F \omega E_i}
\end{equation}
where E$_f$ is the energy distribution with Fano noise added, R$_n$(0) is the normally distributed random number with mean 0 and variance 1, F is the Fano factor = 0.115 and $\omega$ is the average energy required to produce an electron-hole pair in Si= 3.65 eV. A mono-energetic spectrum with a mean energy at 1.25\,keV, incorporated with Fano noise, is simulated at normal incidence to the CCD. Figure~\ref{fig:source_noise_spectrum} (left) shows the histogram of the incident photon spectrum. Each photon is simulated to hit a random pixel location on the CCD. The depth at which each photon interaction occur inside the CCD is determined from:
\begin{equation}
z_0 = -\frac{1}{\mu(E)} ln(R_u)
\end{equation}
where  $\mu$(E) is the linear mass absorption coefficient of Si at photon energy E, R$_u$ is a random number with uniform distribution. Photons interacting at depths ($z_0$) beyond the  substrate thickness are considered to be lost in the simulation. The absorbed photon produces a charge cloud, which is assumed to be spherical with a Gaussian distributed charge density.  The 1-$\sigma$ radius of the initial spherical charge cloud is calculated using \citep{Townsley2002}:
\begin{equation}
    r_i \simeq 0.0062~E_{f}^{1.75} (in ~\mu m)
\end{equation} 
where E$_{f}$ is the photon energy (in keV). The charge cloud drifts under the influence of electric field and undergo random thermal motions resulting in the expansion of the charge cloud radius determined by \citep{hopkinson1987}. The typical final mean-square charge cloud radius for a photon in the {\magixs} energy range lies around $\sim$ 2 to 5 $\mu$m. The amount of charge collected in each pixel for a given event is computed by integrating the Gaussian distributed charge cloud by using Equation 9 from \citep{Athiray2015}. Further, we add random noise to the charge collected in each pixel, which is the residual of measured average CCD darks. The synthetic X-ray image simulated with source photons and noise events is shown in Figure~\ref{fig:source_noise_spectrum} (right). We use this as our test case to reconstruct the incident photon energy and determine the number of X-ray hits on the CCD. We emphasize that photons travelling beyond the depletion depth and photons that hit channel stops are considered to be lost and hence are not included in the event selection. A more detailed charge transport simulation with interactions at different layers on the CCD would help to understand the spectral response \citep[e.g.][]{Townsley2002, Athiray2015, Godet2009,Frank2004}, which is beyond the scope of our current investigation.

\begin{deluxetable}{c c}
\tablecaption{Physical parameters for the simulated photon interaction and charge propagation in the {\magixs} CCD. \label{tab:params}}
\tablehead{
\colhead{Parameters} & \colhead{Values} }
\startdata
Mean photon energy & 1.25 keV \\
Dopant concentration (N$_a$) & 4 $\times$ 10$^{12}$ cm$^{-3}$\\
Temperature & 203 K \\
CCD dimensions & 2 K $\times$ 1K pixels \\
Pixel size & 15 $\mu$m\\
\enddata
\end{deluxetable}

{}

\begin{figure}[h]
    \centering
    \includegraphics[width=0.5\linewidth ]{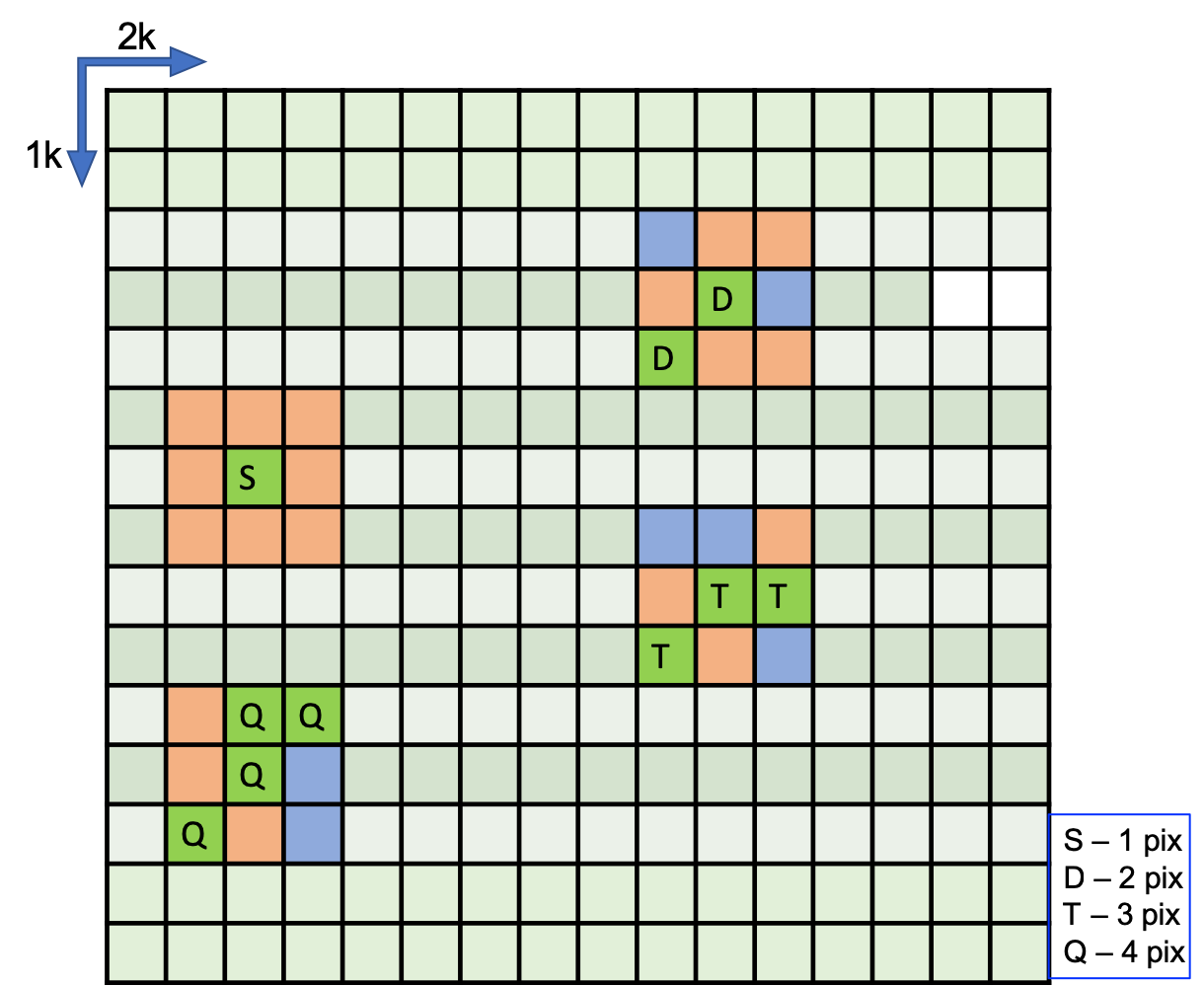}
    \caption{Representation of single and multipixel events in the CCD and event classification. The labeled Green boxes represent pixels with values greater than the baseline threshold (3-$\sigma_{noise}$). Peach color boxes represent the search grid around the bright pixels. Blue color boxes represent pixels within the search grid that have a charge deposit greater than zero and less than baseline threshold, which are summed up with pixels that satisfy the criteria of baseline threshold to reconstruct the photon energy.}
    \label{fig:event_selection}
\end{figure}

\subsection{Event Characterization}
For a typical charge cloud radius of $\sim $5 $\mu$m, photons interacting near to the pixel center register 
as a single event.  Depending on the location of interaction away from the pixel center, events can be registered in more than one pixel. Based on the pixel size and typical charge cloud size we deduce a search grid  $\pm$ 1 pixel around the center hit pixel would be suffice for event characterization. Reconstructing the total energy deposited from a charge shared event by summing the signal over many pixels is further complicated by the detector readout noise. Typically, most of the X-ray imaging instruments using CCDs implement event grading methods to classify pixel patterns.  In this approach, a `baseline' threshold is first applied to identify pixels above certain intensity as a valid X-ray event. We have adopted 3~$\times$~$\sigma_{noise}$ as our {\it baseline threshold} ($\sim$ 0.12 keV) and marked pixels with intensity above this value to be an X-ray event; $\sigma_{noise}$ is the RMS readout noise. Starting with the brightest pixel, we search for additional neighbor hits or {\it `split' events} registered within $\pm$\, 1 pixel from the center bright pixel. This scanning width implies a search over a grid of 3~$\times$~3 pixels. We then classify events based on the number of pixels that are above the baseline threshold within each grid as `1 pix',`2 pix' `3 pix', `4 pix', etc. To recover the total energy deposited and ensure charge is not lost from the shared events, all the pixels with charge deposit greater than zero within each grid are summed up. A schematic representation of the event classification is shown in Figure~\ref{fig:event_selection}. Pixels that satisfy baseline threshold are marked as green boxes, the search grid (3 $\times$ 3 pixels) is shown in peach boxes, blue boxes represent pixels within the search grid that have a charge deposit greater than zero and less than baseline threshold, and labels (`S', `D', `T' and `Q') denote event classification in Figure~\ref{fig:event_selection}. Thresholds are carefully chosen so that noise events do not influence the counting of `real' X-ray photon events. Complications arising from charge transfer inefficiency are excluded in the model for simplicity.

\begin{figure}[h]
 
    \includegraphics[width=0.5\linewidth]{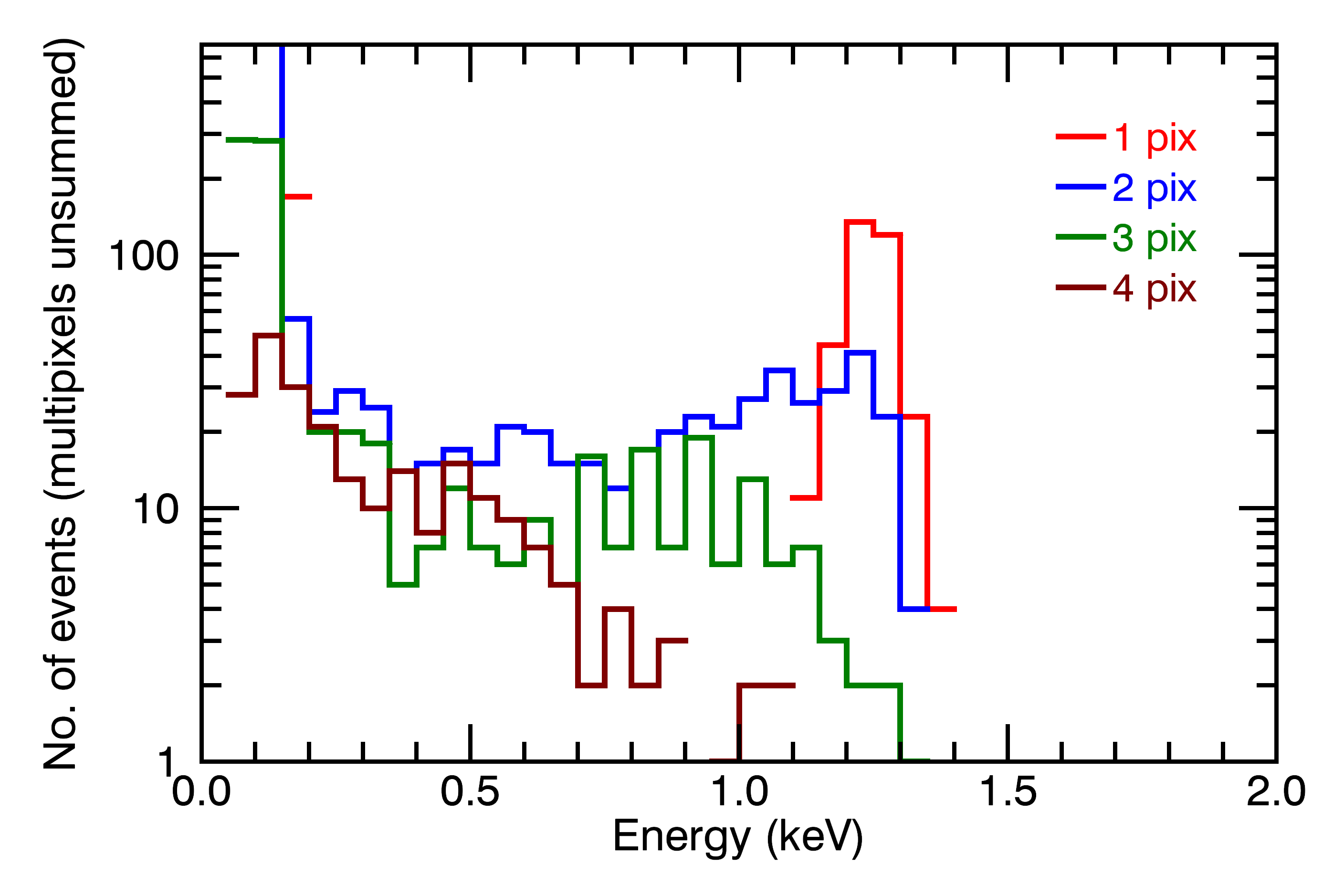}
    \includegraphics[width=0.5\linewidth]{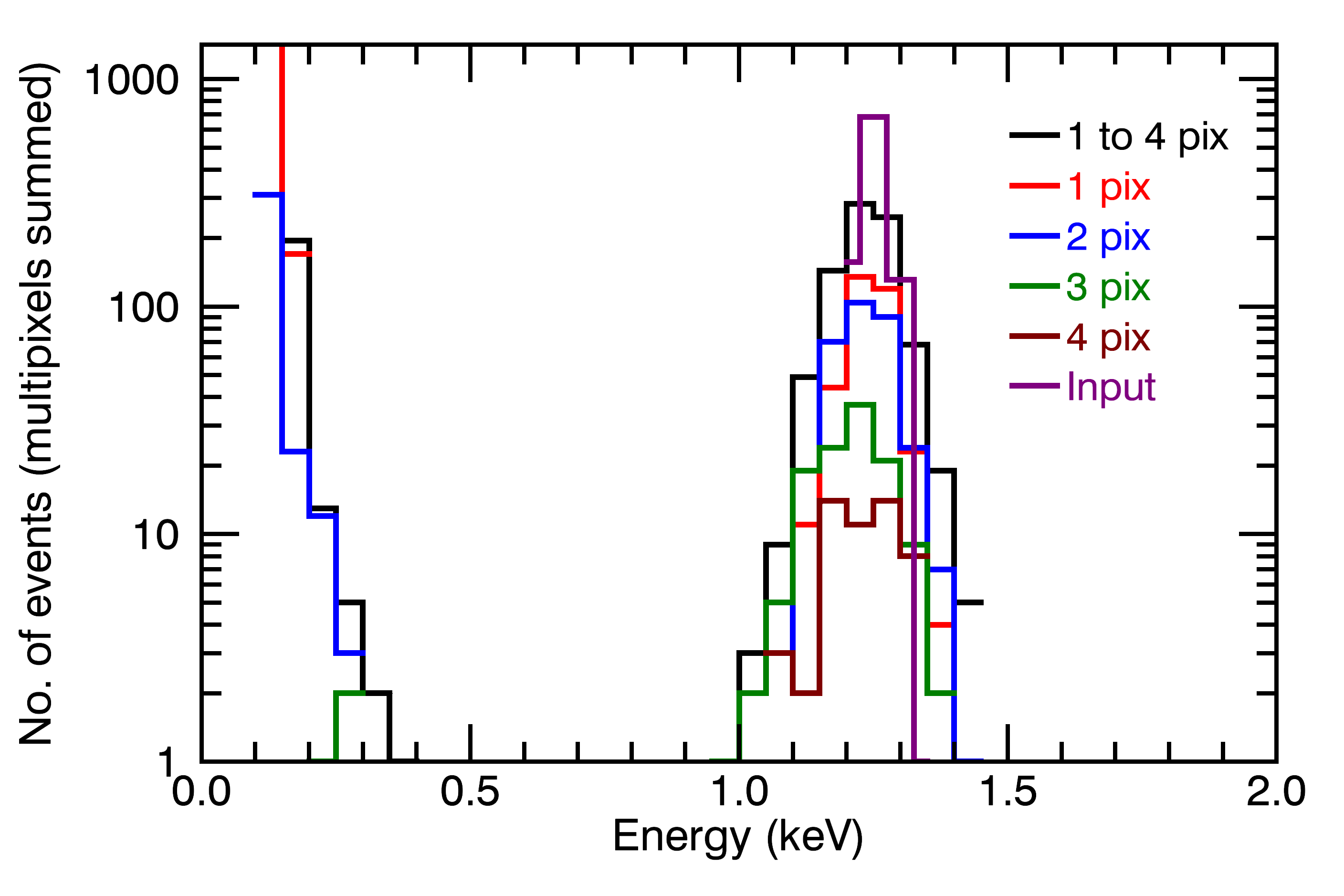}
    \caption{The histogram of simulated events registered on the CCD satisfying the event selection criteria and classification.  (Left) Histogram of the observed events classified based on the number of pixels above baseline threshold in each search grid, without recovering energy from the charge shared events. It is evident that significant fraction of multipixel events are close to detector noise levels. (Right) Histogram of events
    processed through energy recovery by summing up pixels with charge deposits greater than zero in each search grid. The energy reconstructed from the multipixel events using the event selection algorithm matches well with the incident energy.}
    \label{fig:classified_recovered_pixel_hit_spectrum}
\end{figure}{}
Figure~\ref{fig:classified_recovered_pixel_hit_spectrum} (left) shows the histogram of energy measured in individual pixels that form single or multipixel events. Though there are a few single pixel events detected (black line), a significant fraction of events undergo charge sharing and are close to detector noise levels. This measurement implies that proper energy reconstruction of the incident photon is critical to precisely determine the incident photon flux. It also conveys that the charge induced by 1.25\,keV photons can be distributed over as many as 4-pixels. The energy reconstructed spectrum is shown in Figure~\ref{fig:classified_recovered_pixel_hit_spectrum} (right) along with the simulated incident spectrum overplotted. This demonstrates the energy reconstructed from multipixel events using the event selection algorithm described above matches well with the incident photon energy. 

\begin{figure}
    \centering
    \includegraphics[width=0.5\linewidth]{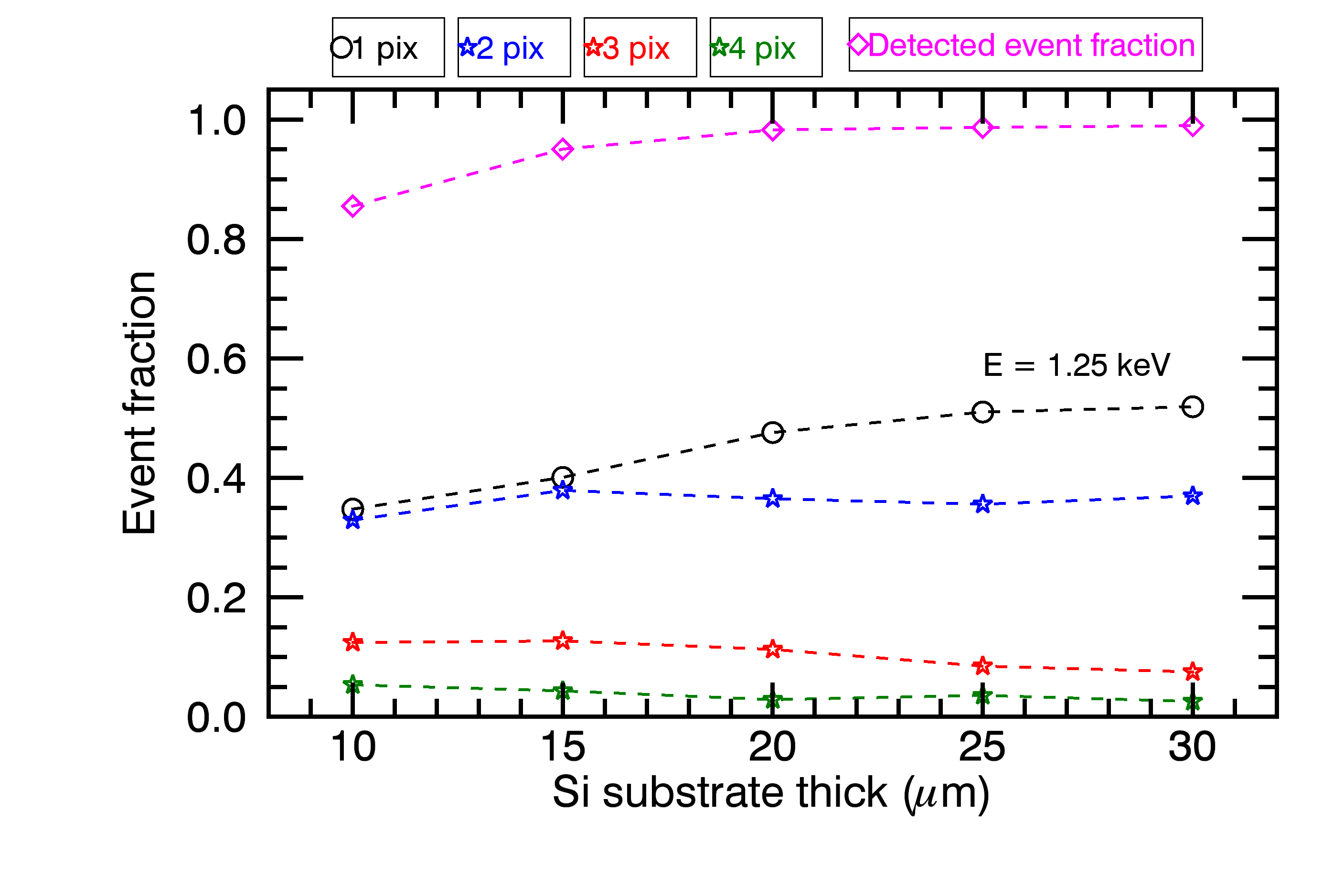}
    \caption{Si substrate thickness Vs fraction of single and multipixel events. The fraction of single pixel events systematically increase with Si thickness. This implies increase in collection efficiency, which is shown as increase in the total detected event fraction. However, the fraction of multipixel events are less sensitive to substrate thickness.}
    \label{fig:depth_dependence}
\end{figure}{}

Further, we also tried to understand the sensitivity of Si substrate thickness to result in single and multipixel events, within {\magixs}'s wavelength range (6 - 24 {\AA}). For this, we performed a simulation with different substrate thicknesses ranging from 10 $\mu$m to 30 $\mu$m, retaining all other parameters such as the pixel geometry etc fixed. The simulation assumes the substrate thicknesses to be a {\it field-region}, where there is a strong electric field to drift the charge cloud. The fraction of single pixel and multipixel events varying as a function of substrate thickness for photon energy E = 1.25 keV is shown in Figure \ref{fig:depth_dependence}. This clearly indicates the fraction of single pixel events increases with substrate thickness, which means more collection efficiency as seen from the detected event fraction. It also suggests the fraction of multipixel events is much less sensitive to substrate thickness. This result could be explained as the substrate thickness increases, the active volume of the detector i.e. region with electric field increases, which improves the probability of photon detection from a relatively greater depth, thereby increasing the overall quantum efficiency of the detector. The strong electric field created while increasing the active volume of the detector yields a relatively small cloud radius at these photon energies, which improves the fraction of single-pixel events, but  does not significantly affect the fraction of multipixel events arising from the weak electric field regions. We interpret the observed charge shared events as mainly due to photon interactions near the pixel boundaries, and therefore describe this as a 3D geometrical effect. 
 
\section{Results from Experimental Data}
\label{sec:results}
\subsection{Applying Event Detection to CCD Data}
\begin{figure}[h]
        \includegraphics[width=0.5\linewidth]{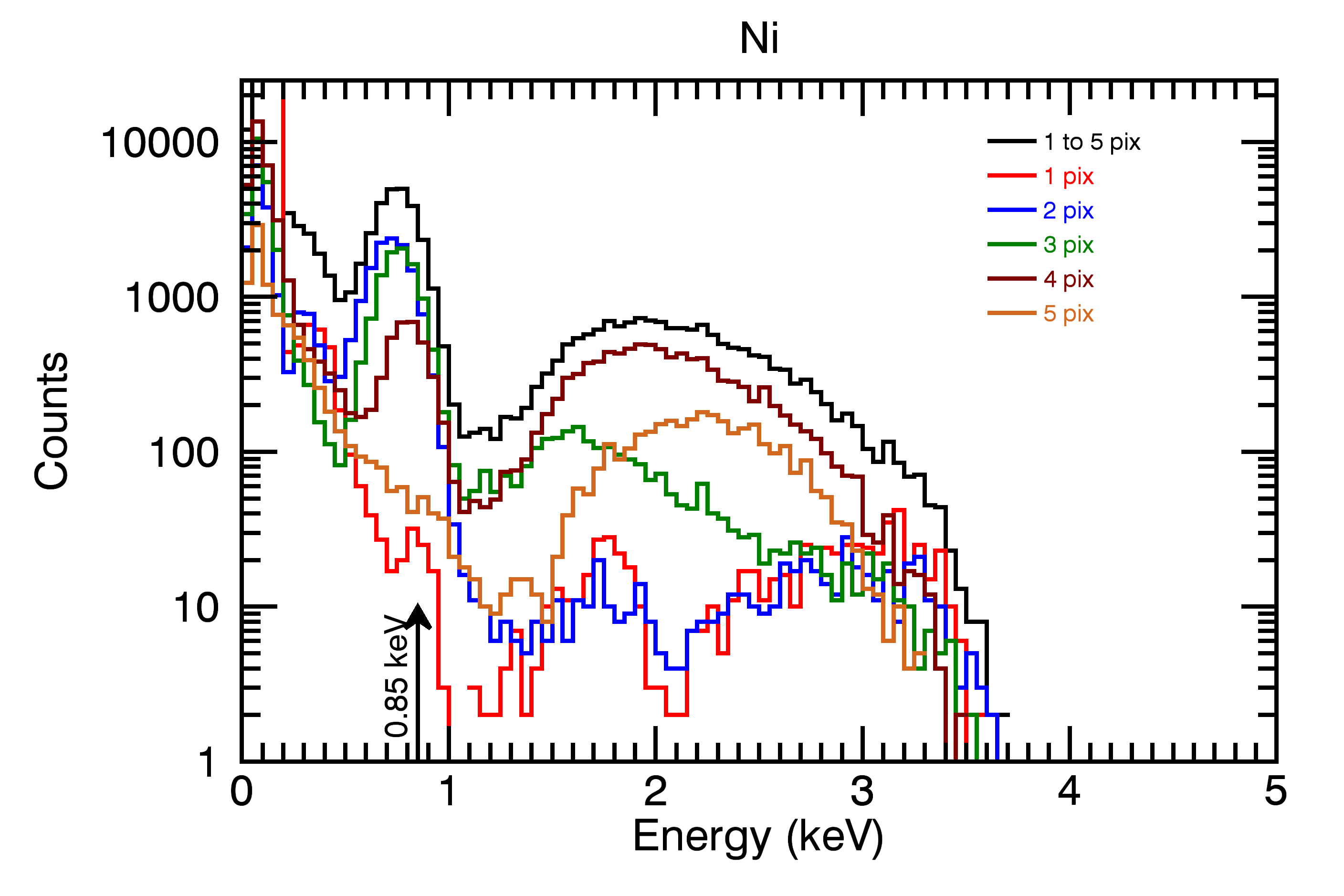}
        \includegraphics[width=0.5\linewidth]{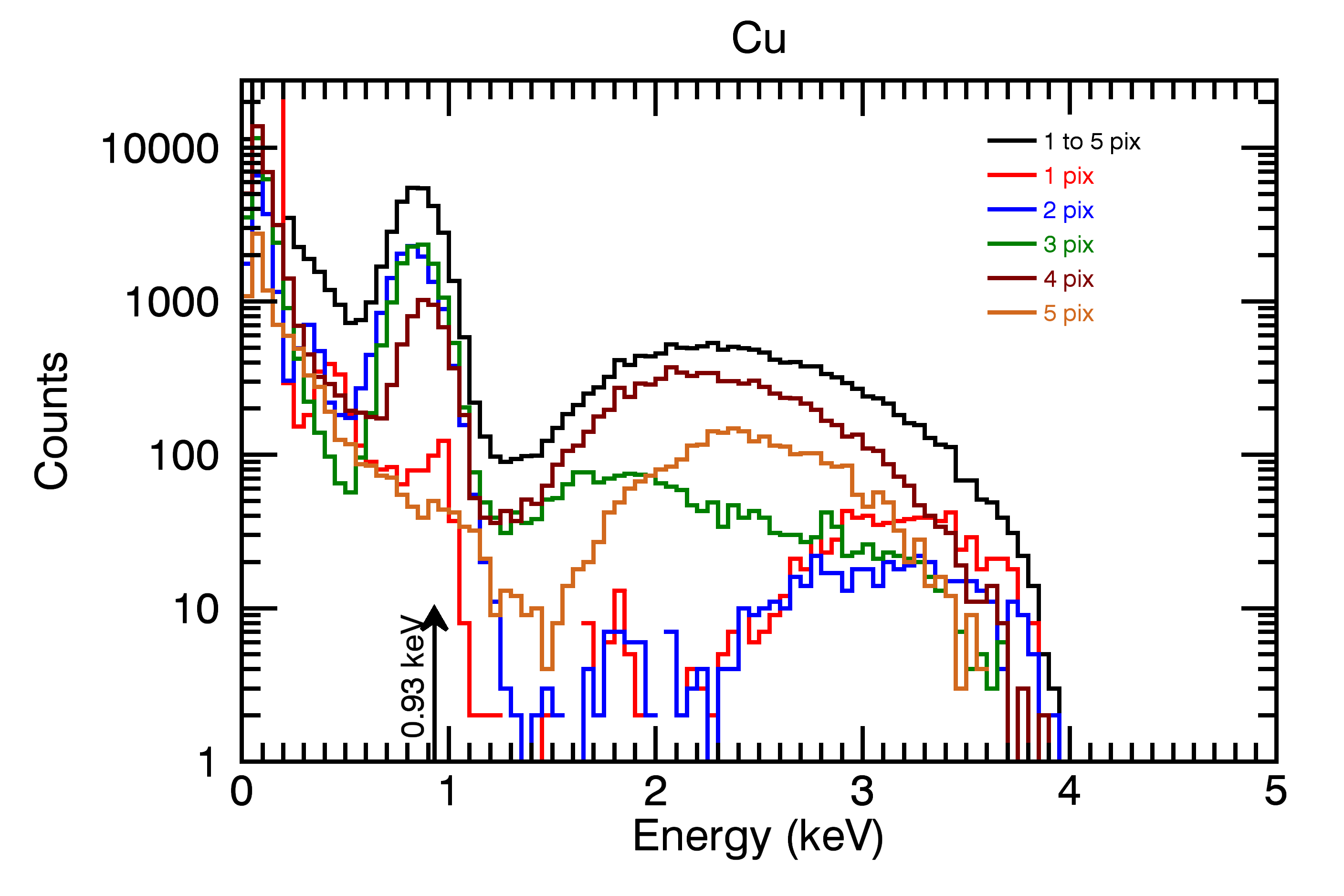}       \includegraphics[width=0.5\linewidth]{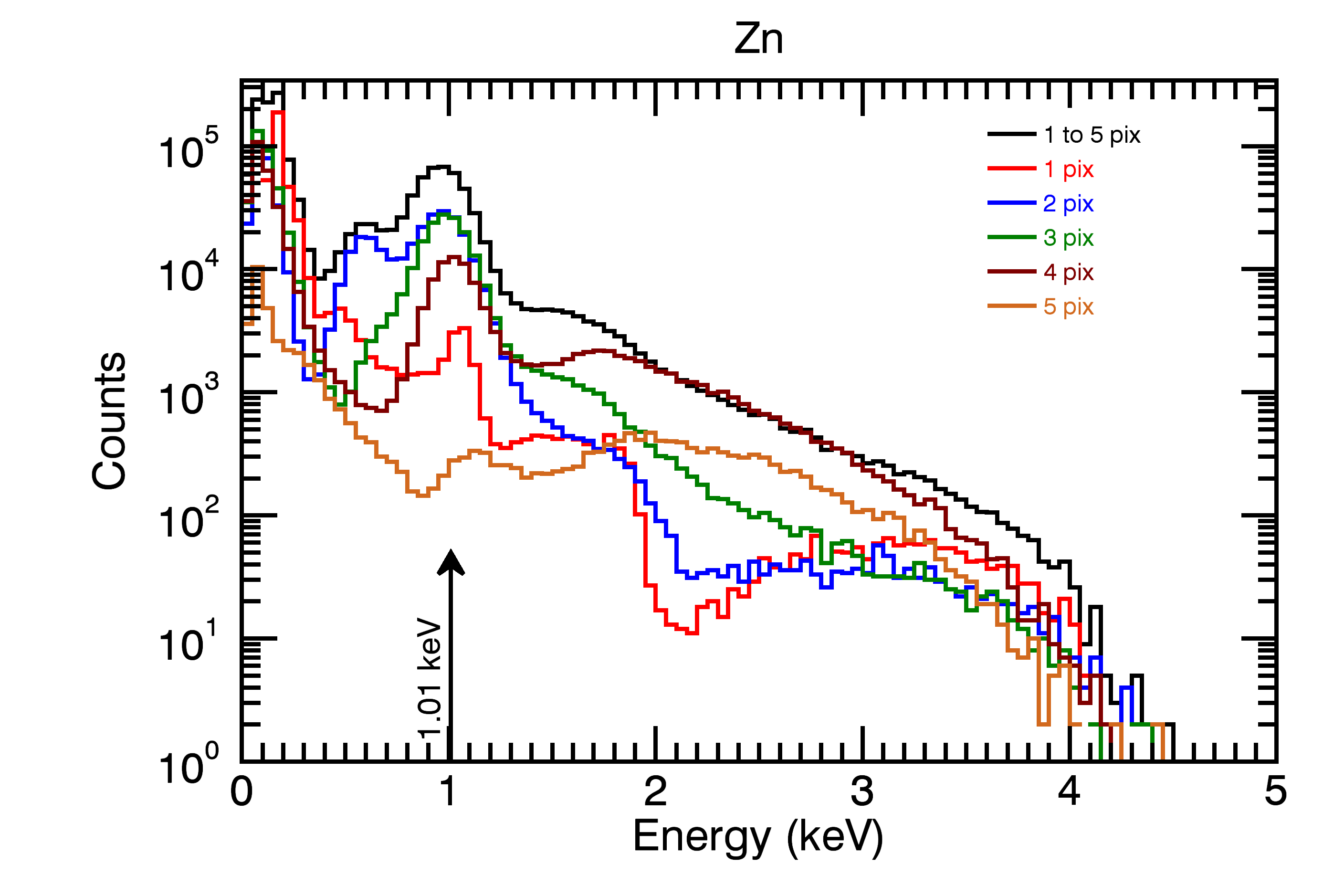}
        \includegraphics[width=0.5\linewidth]{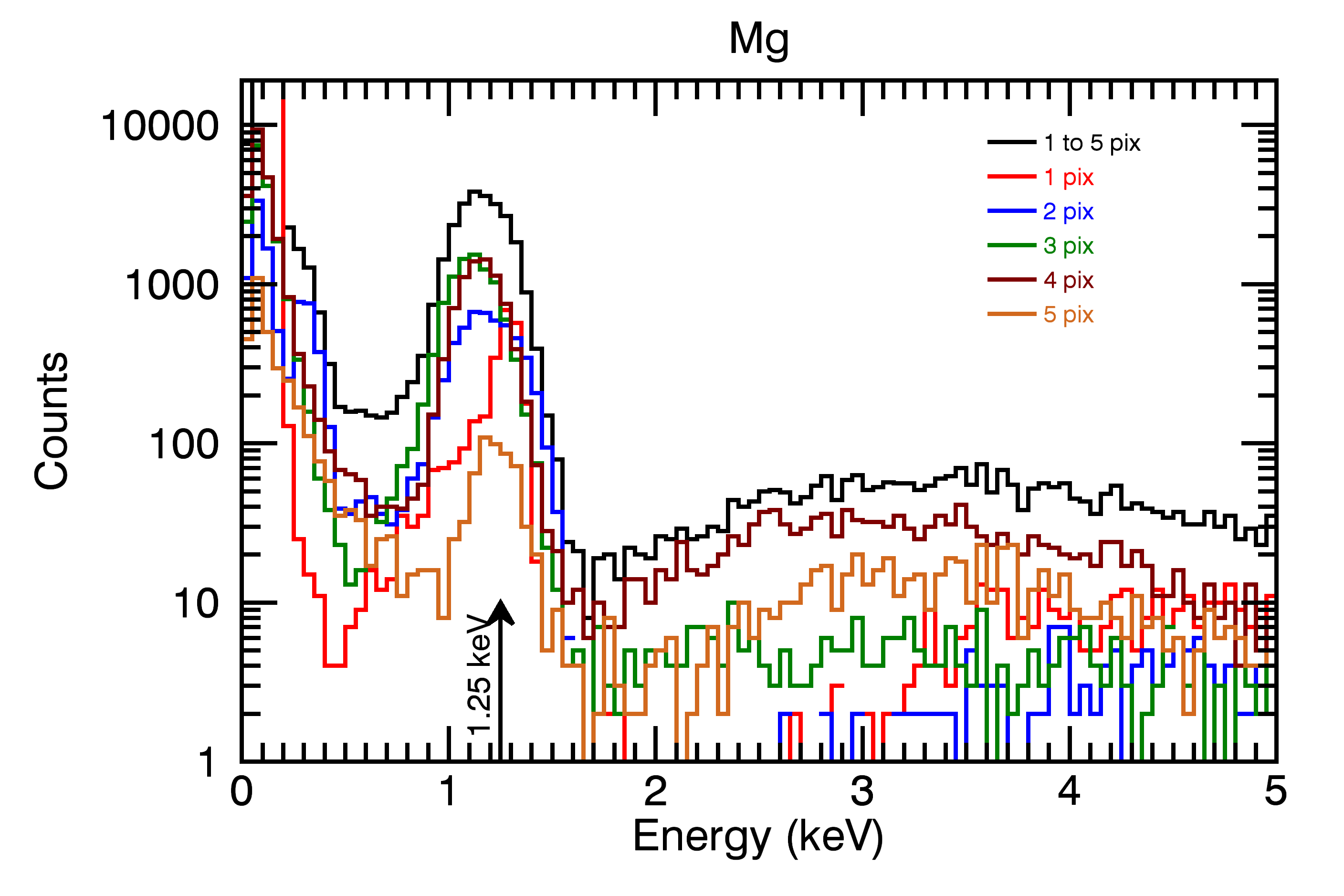}
    \caption{Histograms of X-ray events reconstructed for targets Mg-K, Ni-L, Cu-L, and Zn-L. The photon energy derived from single pixel events is clearly conspicuous at the respective line energies of the target. For multipixel events the reconstructed energy is systematically less than the line energy and exhibits a broad spectral response, consistent with \citep{Eric2019}.}
    \label{fig:expt_classified_recovered_pixel_hit_spectrum}
\end{figure}

We processed the experimental data through the same event selection logic used in the simulation  and determined the incident photon flux for different targets listed in Table~\ref{tab:calib_targets}. The best comparison of simulations and real data is the distribution of multipixel events. A sample of distributions of 1 pix, 2 pix, 3 pix, 4 pix, and 5 pix events recovered from the event processing for different targets is shown in Figure~\ref{fig:expt_classified_recovered_pixel_hit_spectrum}. The results indicate a significant fraction of photon hits lead to multipixel events, which is consistent with our simulation result. However, we observe that multipixel events are more pronounced in real data as compared to simulations. Energy reconstruction from single pixel events matches well with the incident photon energy, which is marked with a vertical arrow in Figure~\ref{fig:expt_classified_recovered_pixel_hit_spectrum}. However, summing up of pixels from all of the charge shared events does not lead to proper reconstruction of the incident photon energy. The reconstructed energy from  multipixel events show consistently less than the incident photon energy, which is not the case in our simulation. This discrepancy could be due to incorrect modeling of charge diffusion and/or lack of charge loss mechanism in our `ideal' simulation. Also additional effects, such as pixel non-uniformity, charge transfer inefficiency, and distortion in electric field distributions, could 
account for loss in the total energy collected from multipixel events. It  has already been shown that energy reconstruction of X-rays below 2\,keV is highly complicated for detectors with large depletion and small pixel sizes \citep{Eric2019}, which is consistent with our observations.

{}

\subsection{Comparison of spectra with BND}
Finally, we compare the spectra from the processed CCD data described above with the spectra obtained from the BND. Note that these two detectors are completely different in terms of noise characteristics, spectral response function, and quantum efficiency, which are energy dependent parameters. These distinctions are acceptable since the spectral comparisons are semi-qualitative considering our objectives 1) to determine and compare the  X-ray intensity at target's line energy and 2) to look for similarities in the spectral profile, including the line to continuum. 

\begin{figure}
    \includegraphics[width=0.5\linewidth]{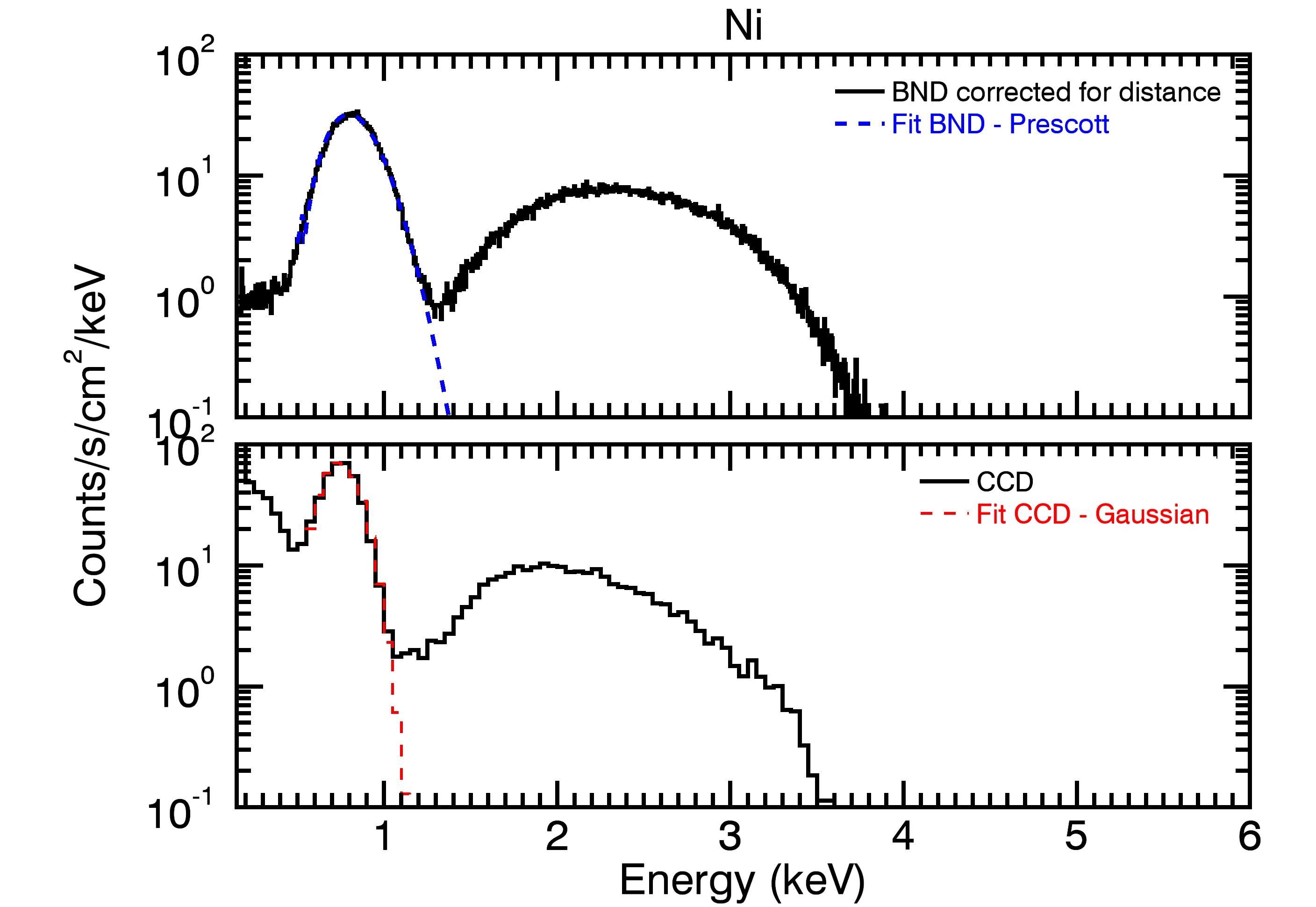}
    \includegraphics[width=0.5\linewidth]{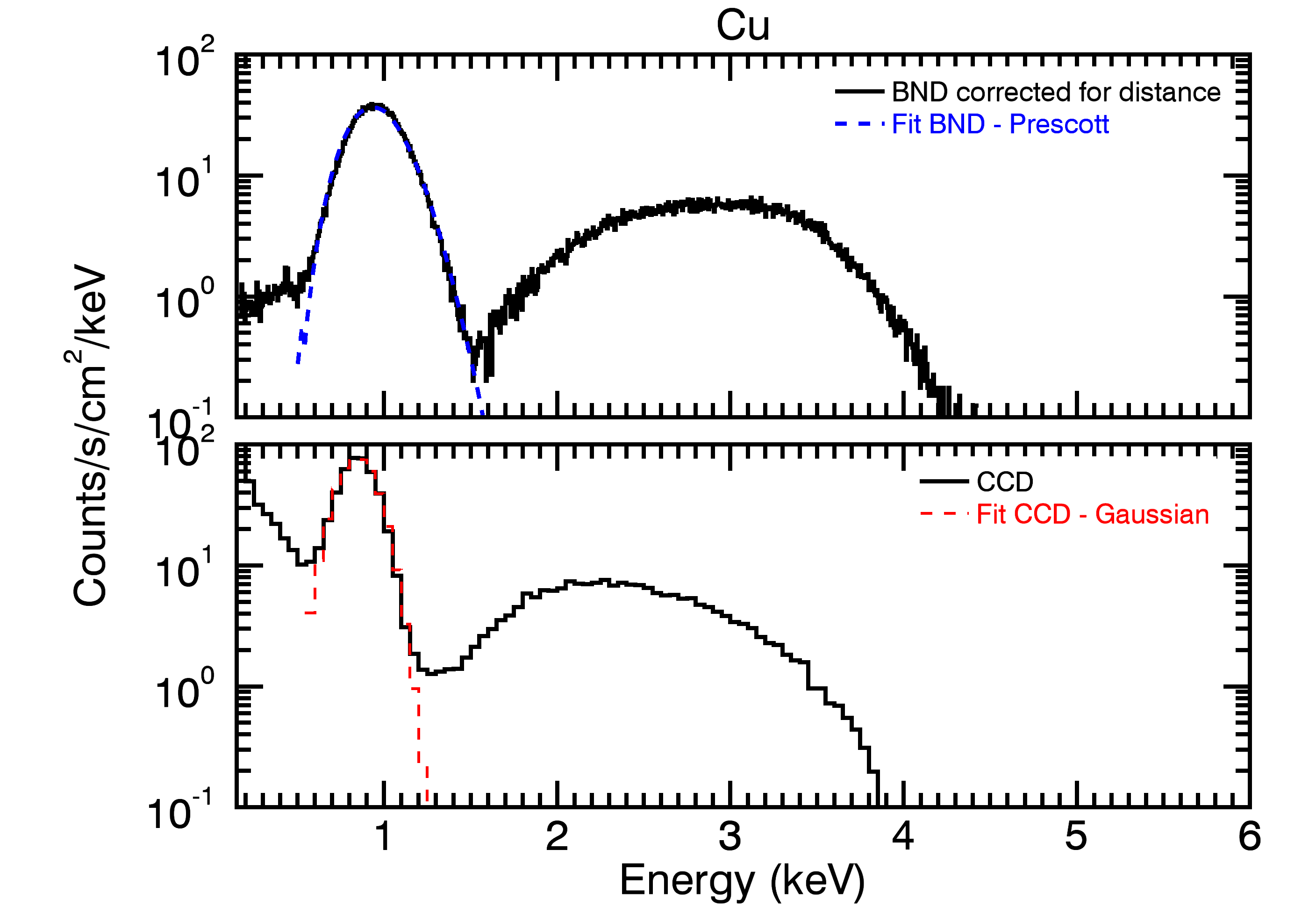}
    \includegraphics[width=0.5\linewidth]{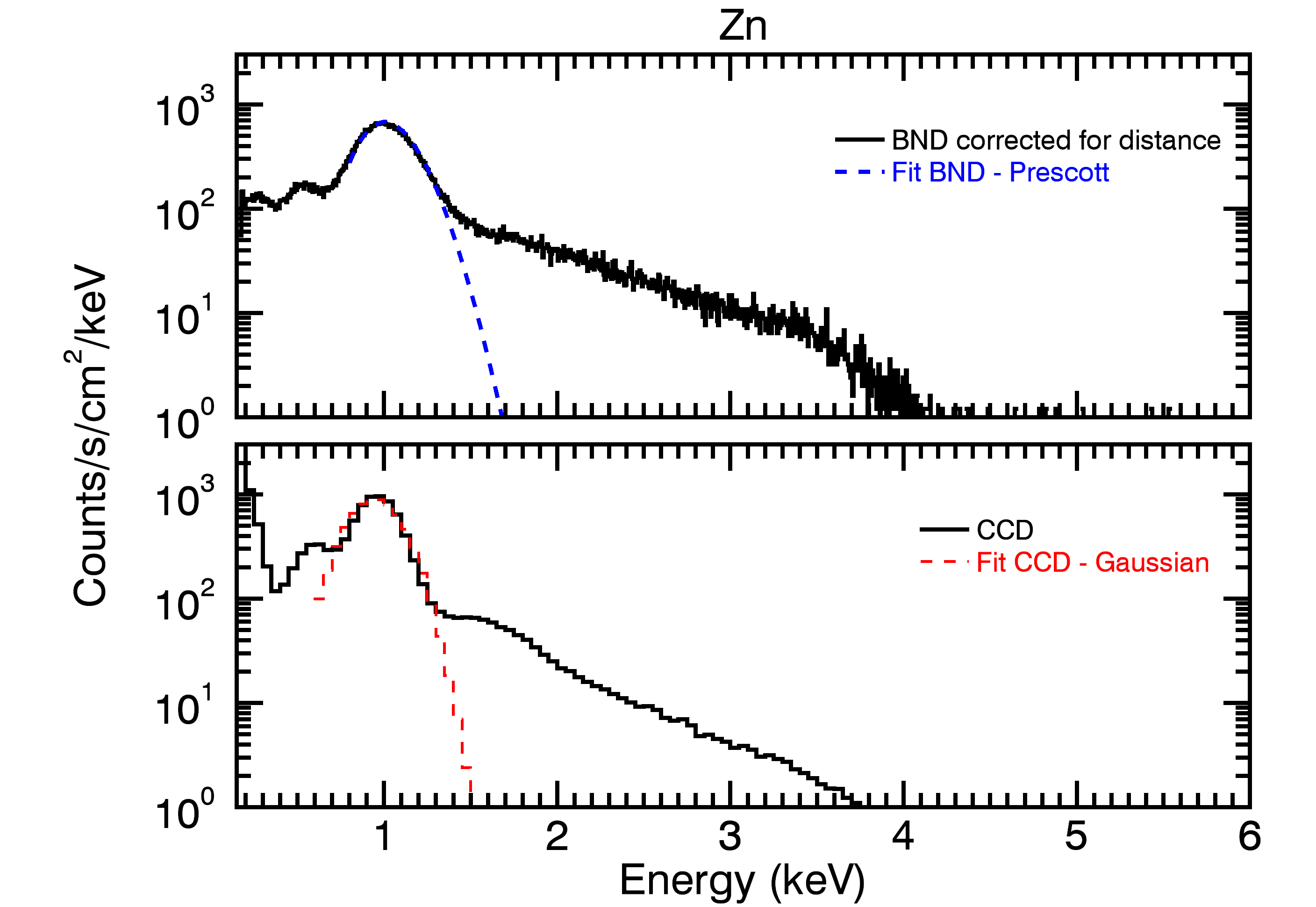}
    \includegraphics[width=0.5\linewidth]{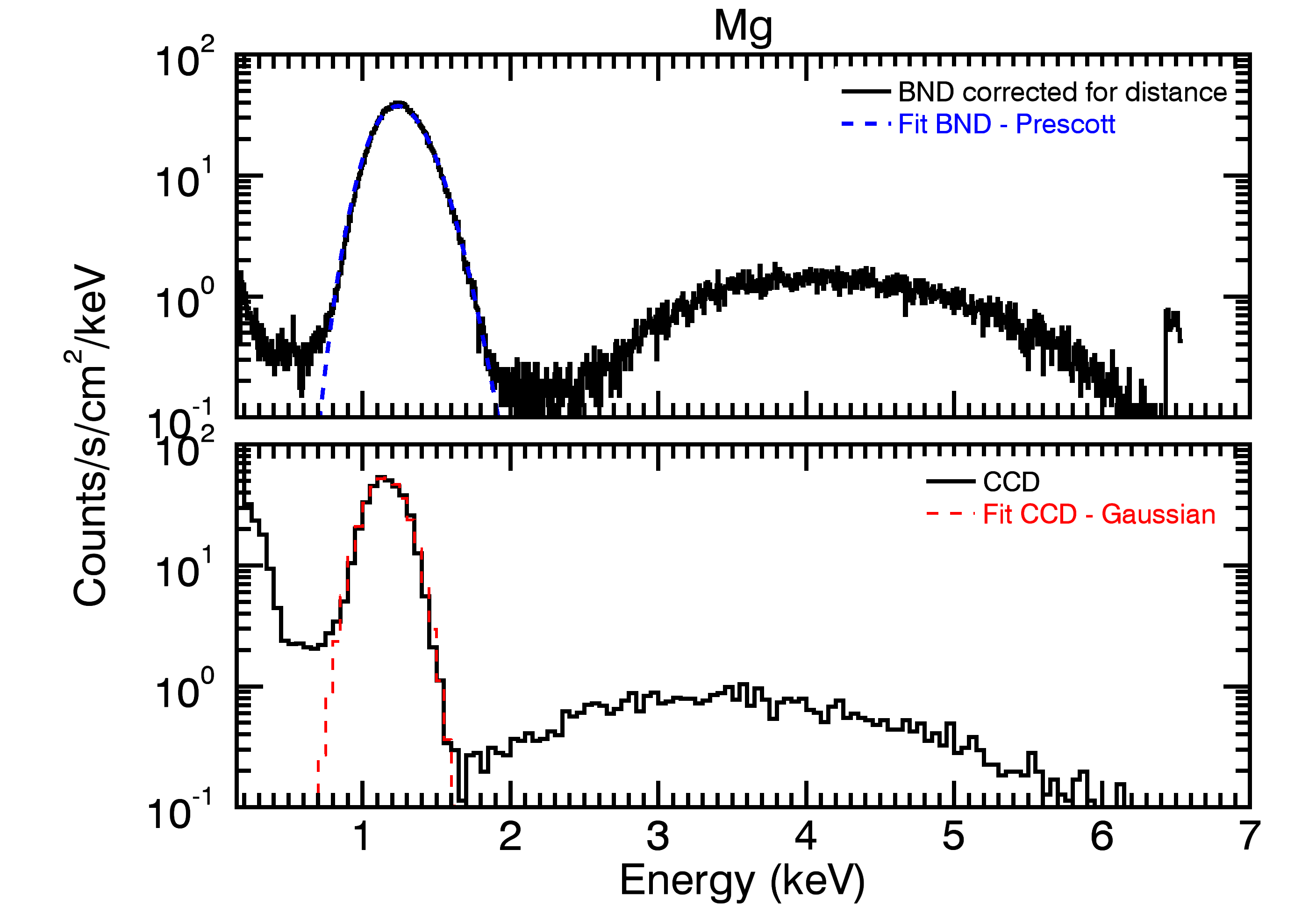}
    \caption{Comparison of spectra from different targets with the histograms of X-ray events reconstructed for the corresponding targets. The line peaks are modeled with a Gaussian function for the CCD spectra along with a Prescott function for the BND spectra to determine the intensities at respective target line energies, which are compared in Table~\ref{tab:lineflux}.}
    \label{fig:ccd_fpc_spectracomparison}
\end{figure}

 As the detectors are placed at different distances from the source, a scale factor is applied to the BND data to project the spectra at the appropriate CCD distance.  The comparison of spectra from the CCD and BND (after taking into account respective integration times, and collection area) is shown in Figure~\ref{fig:ccd_fpc_spectracomparison}.  Spectra from targets Ni, Cu, and Mg are taken with respective filters, which preferentially transmit the target's characteristic X-rays and suppress the continuum. In contrast, spectra from the Zn target without a filter (Figure~\ref{fig:ccd_fpc_spectracomparison} lower left) clearly shows the line riding over a dominant bremsstrahlung continuum.  It is evident the overall spectral profile for different targets from both detectors looks similar. However, there are discrepancies that are clearly visible between the spectra, especially at the high energy tail of the line and the continuum.  We ascribe the discrepancy in the continuum between the detectors chiefly arising from the inherent spectral response of the detector.

It has been already shown by \citet{Auerhammer98} that the spectral response of the BND exhibits a skewed peak and an energy dependent low energy shelf, as measured at several discrete monochromatic energies. Further studies indicate that the skewness of the peak can be best modeled by a Prescott function \citep{Prescott1963}, which has a heritage of application for proportional counters \citep{Budtz1995}.  Other factors, such as a small fraction of pile-up events or energy-dependent spectral response function of the detectors, can introduce additional second order effects that cause discrepancies between the observed spectra from the two detectors.  We modeled the line peak in the CCD spectra with a Gaussian function along with a Prescott function for the BND spectra to determine the intensity at target's line energy and then compare the X-ray flux as given in Table~\ref{tab:lineflux}. We used the quantum efficiency of the BND \citep{Martin1997} and the quantum efficiency of the CCD determined from the simulation to convert the line fluxes from counts to respective photon units. The quantum efficiency of the CCD obtained from the simulation closely agrees with the published quantum efficiency for the back-illuminated, astro-processed CCDs from e2V Technologies Ltd. \citep{Moody2017}, which are demonstrated to be reliable and consistent in the {\magixs} energy range. The error in the CCD flux is derived from the quadrature sum of statistical uncertainty and systematic uncertainty from the CCD gain variations. The error in BND flux is determined from the quadrature sum of statistical uncertainty and from Section 3, we demonstrated the confidence on the event selection method for CCD in the recovery of photon energy accurately and hence the incident flux. Therefore, the idea for comparing the BND flux against CCD is to determine the level of uncertainty in the BND flux. From Table \ref{tab:lineflux}, we observe the incident X-ray flux determined from the CCD agrees within 20\% of the incident flux measured from the BND.

\begin{deluxetable}{c c c c c c c c}
\tablecaption{Incident X-ray flux determined from the BND and the CCD for different calibration targets at  respective target's line energy \label{tab:lineflux} .}
\tablehead{
\colhead{Target} &\colhead{Line energy}& \colhead{Voltage}& \colhead{Current}& \colhead{Filter} & \colhead{BND flux} & \colhead{CCD flux} & \colhead{Percentage}\\
\colhead{}&\colhead{keV}&\colhead{kV}& \colhead{mA}&\colhead{}&\colhead{Ph/s/cm$^2$/mA} & \colhead{Ph/s/cm$^2$/mA}&\colhead{difference}}
\startdata
Ni-L& 0.85& 3.40 & 5.8 & Ni-L2 &4.6 $\pm$ 0.16  & 3.72 $\pm$ 0.04 &-19.7\\
Cu-L & 0.93& 3.70 & 7.7 & Cu-L2&3.4 $\pm$ 0.12 &3.08 $\pm$ 0.04 & -8.4\\
Zn-L & 1.01& 4.00 & 3.0 & No filter &145.0 $\pm$ 5.1 &124.78 $\pm$1.26  & -14.0\\
Mg-K & 1.25& 6.50 & 1.5 &Mg-K2&14.4 $\pm$ 0.56 &12.67 $\pm$ 0.15 & -11.4\\
    \enddata
 \tablecomments{The BND fluxes listed here are determined after correcting for distance from the CCD.}
\end{deluxetable}

{}

\section{Summary}
\label{sec:summary}
In this paper, we presented the experiments and data processing techniques implemented to verify the absolute number of photons entering the instrument from the X-ray source at the XRCF. This characterization of the source throughput is critical for its utilization in the alignment and calibration activities of the {\magixs} experiment, as measured by a CCD detector. Simultaneous measurement of the incident spectra was obtained using a BND that was calibrated in 1998, which is used to cross-calibrate the incident X-ray flux. Using Monte-Carlo simulations of a CCD detector with simple approximations, we first created synthetic multipixel events with realistic detector noise and optimized our event selection algorithm.  We find that knowledge of pixel-based noise sources is critical for soft X-ray photons to achieve proper energy reconstruction and absolute photon counting. Applying the algorithm to the simulated data, we verified that proper energy reconstruction could be achieved and demonstrated photon counting.  Furthermore, we studied the dependence of event distribution with Si substrate thickness by performing simulations with different Si substrate thicknesses modeled as {\it field-region}. Our findings indicate that the fraction of single pixel events increase with substrate thickness, while the fraction of multipixel events appears to be less sensitive to substrate thickness and mainly depends on the interactions near to pixel boundaries.

We then applied the event selection algorithm on the real experiment data from the CCD and classified multipixel events. The observed multipixel events are more pronounced in real data than our simulations. We find that the energy reconstructed from multipixel events systematically appear at lower energy than the incident photon energy. We then compared spectra from different targets obtained from both the BND and the CCD after taking into account their respective distances, integration time, and quantum efficiency.  Though the overall spectral profile from both detectors showed similarities, discrepancies are noticeable in the spectral redistribution function between the BND and the CCD. This disparity warrants advanced spectral modeling including a detailed charge transport simulation, which are beyond the purview of our current investigation. With the confidence gained from the CCD event selection method for precise photon energy and flux estimation, we compared the intensity of different targets, at the respective line energies, observed by both detectors. The measured incident photon flux from both the BND and the CCD show agreement to within 20\%. This result of validating or cross-calibrating the incident photon flux measured simultaneously by the BND will enable radiometric calibration for the {\magixs} instrument and for any future space instrument characterization.

\acknowledgements
 P. S. Athiray`s research is supported by an appointment to the NASA Postdoctoral Program at the Marshall Space Flight Center, administrated by Universities Space Research Association under contract with NASA.  The {\magixs} instrument team is supported by the NASA Low Cost Access to Space program. The authors would like to thank Dr. Jeffery Kolodziejczak at MSFC for helpful discussions and insights in preparing this manuscript. The authors gratefully acknowledge the many people at XRCF who have contributed to the testing. The authors appreciate helpful and insightful comments and suggestions from an anonymous referee.

\bibliography{sample63,solar,references}{}
\bibliographystyle{aasjournal}



\end{document}